**Title:** Localized Interlayer Excitons in MoSe$_2$-WSe$_2$ Heterostructures without a Moiré Potential

**Author Names:** Fateme Mahdikhanysarvejahany[1*], Daniel N. Shanks[1*], Mathew Klein[1], Qian Wang[2], Michael R. Koehler[3], David G. Mandrus[4-6], Takashi Taniguchi[7], Kenji Watanabe[8], Oliver L.A. Monti[1,9], Brian J. LeRoy[1], and John R. Schaibley[1]

**Author Addresses:**
[1]Department of Physics, University of Arizona, Tucson, Arizona 85721, USA
[2]Guangdong Provincial Key Laboratory of Quantum Metrology and Sensing & School of Physics and Astronomy, Sun Yat-Sen University (Zhuhai Campus), Zhuhai 519082, China
[3]IAMM Diffraction Facility, Institute for Advanced Materials and Manufacturing, University of Tennessee, Knoxville, TN 37920
[4]Department of Materials Science and Engineering, University of Tennessee, Knoxville, Tennessee 37996, USA
[5]Materials Science and Technology Division, Oak Ridge National Laboratory, Oak Ridge, Tennessee 37831, USA
[6]Department of Physics and Astronomy, University of Tennessee, Knoxville, Tennessee 37996, USA
[7]International Center for Materials Nanoarchitectonics, National Institute for Materials Science, 1-1 Namiki, Tsukuba 305-0044, Japan
[8]Research Center for Functional Materials, National Institute for Materials Science, 1-1 Namiki, Tsukuba 305-0044, Japan
[9]Department of Chemistry and Biochemistry, University of Arizona, Tucson, Arizona 85721, USA

**Corresponding Author:** John Schaibley, johnschaibley@email.arizona.edu

**Abstract:**

Trapped interlayer excitons (IXs) in MoSe$_2$-WSe$_2$ heterobilayers have generated interest for use as single quantum emitter arrays and as an opportunity to study moiré physics in transition metal dichalcogenide (TMD) heterostructures[1,2,3]. IXs are spatially indirectly excitons comprised of an electron in the MoSe$_2$ layer bound to a hole in the WSe$_2$ layer[4,5]. Previous reports of spectrally narrow (<1 meV) photoluminescence (PL) emission lines at low temperature have been attributed to IXs localized by the moiré potential between the TMD layers[6,4,7,8,9,10]. Here, we show that spectrally narrow IX PL lines are present even when the moiré potential is suppressed by inserting a bilayer hexagonal boron nitride (hBN) spacer between the TMD layers. We directly compare the doping, electric field, magnetic field, and temperature dependence of IXs in a directly contacted MoSe$_2$-WSe$_2$ region to those in a region separated by bilayer hBN. Our results show that the localization potential resulting in the narrow PL lines is independent of the moiré potential, and instead likely due to extrinsic effects such as nanobubbles or defects. We show that while the doping, electric field, and temperature dependence of the narrow IX lines is similar for both regions, their excitonic g-factors have opposite signs, indicating that the IXs in the directly contacted region are trapped by both moiré and extrinsic localization potentials.

**Main Text:**

Localized excitons (coulomb-bound electron-hole pairs) which can serve as single photon emitters have been investigated for decades due to their potential applications in quantum information science and optoelectronics. Recently there has been significant interest in moiré effects in 2D material heterostructures that arise from the in-plane superlattice potential that occurs between two twisted or lattice mismatched layers[1,3,12]. Seyler et al.[1] were the first to report spectrally narrow photoluminescence (PL) arising from interlayer excitons (IXs) in $MoSe_2$-$WSe_2$ heterostructures. The narrow PL lines were observable only at low temperatures (<~15 K) and low (< ~100 nW) excitation power and consisted of an inhomogeneous distribution of narrow lines with a spread of order of 20 meV. The narrow PL lines were attributed to single IXs trapped to moiré potential sites, as evidenced by the spectrally narrow (< 1 meV) PL emission and circularly polarized optical selection rules. There have been numerous papers studying these narrow lines since the initial report including reports of single photon emission, charged IXs, and Coulomb staircase effects[13,14,15]. In this work, we show that the spectrally narrow IX PL lines are still present in $MoSe_2$-hBN-$WSe_2$ heterostructures, where the moiré potential is suppressed by an hBN spacer layer (see Extended Data Fig. 1). We compare the physical behaviors of narrow IX PL emission from non-separated, directly contacted (DC) and hBN-separated sample regions and show that the localization potential resulting in the narrow IX lines is likely due to an extrinsic disorder effect, not the moiré potential.

The sample structure is comprised of a twist-angle aligned $MoSe_2$-$WSe_2$ heterostructure as depicted in Figure 1a (see Methods). In order to probe the dependence of the IX trapping on the moiré potential, we fabricated a device where half of the heterostructure has the TMD layers separated by bilayer hBN and the other half has them in direct contact (with no hBN spacer). The TMD layers are encapsulated in hBN, and a few layer graphene top gate ($V_t$) and graphite back gate ($V_b$) serve to independently control the charge density and external electric field experienced by the TMD layers. We used low temperature confocal PL spectroscopy to measure the spatial dependence of the IX emission photon energy (Figure 1b). In the DC region of the device the IX photon energy is centered around 1.34 eV, consistent with R-type (near 0° degree twist) $MoSe_2$-$WSe_2$ heterostructures[16]. The hBN separated region shows higher energy PL centered around 1.42 eV. This 80 meV increase in energy of the PL is in agreement with the suppression of the moiré potential by the insertion of bilayer hBN (See Extended Data Fig.1). When the confocal pinhole was removed, PL from both regions could be detected when exciting on the higher energy hBN-separated region. Figure 1c shows co- and cross-circularly polarized PL when exciting the hBN-separated region with a 1.72 eV laser and detecting from both regions. Surprisingly, while the DC region contacted shows cross-circularly polarized PL, consistent with previous studies on R-type structures[1,2,16], the hBN-separated region shows mostly co-circularly polarized PL (see Extended

Data Fig. 2 for data from another sample). We note that the co-circularly polarized emission is consistent with the suppression of the moiré potential. The moiré potential traps excitons at the $R_h^X$ stacking site where the chalcogen atoms of the MoSe$_2$ layer are directly above the empty center of the hexagon in the WSe$_2$ layer. When exciting with a $\sigma^+$ laser, most of the IXs will comprise an electron in the K'-valley and a hole in the K-valley. The $C_3$ symmetry of the lowest energy stacking configuration gives rise to the optical selection rule that this site must emit $\sigma^-$ circularly polarized photons. The loss of the cross-circularly polarized PL emission indicates that IXs are no longer trapped at the $R_h^X$ site when a bilayer hBN spacer is inserted. We also note that both the DC and hBN-separated regions show IXs with an inhomogeneous distribution of spectrally narrow (< 1 meV) IX lines demonstrating that they are not due to the moiré potential.

In order to confirm that the narrow PL emission in the hBN separated region originates from IXs, we investigated the electric field and doping dependence which was achieved by applying simultaneous top and back gate voltages. In these measurements, we again excited the hBN separated region and measured the PL from both regions. Figure 2a shows the electric field dependence of the PL while keeping the sample charge neutral[17]. In the DC region, we observe a Stark shift of 0.6 eV/(V/nm) that matches with the known dipole moment of the IX[17,18]. By adding the hBN bilayer between the TMD layers, we increase the separation between the electron and hole and consequently the IX permanent dipole moment will increase by a factor of two that perfectly matches with our finding yielding a 1.2 eV/(V/nm) energy shift, consistent with previous reports on hBN separated IXs[11]. We also explored the IX doping dependence for both sample regions (Figure 2b). We identify three doping regions corresponding to i-electron doped, ii-intrinsic, and iii-hole doped similar to previous reports of charged IXs[13,14,17,19]. The relatively small intrinsic doping range (-0.1 to 0.1×10$^{12}$ cm$^{-2}$) is consistent with a high quality device. We note that in both regions, the IX energy increases with doping which is consistent with previous reports[14,17]. We also note that the DC region shows more prominent fine structure in its doping dependence which was previously attributed to a Coulomb staircase effect[14] (see Extended Data Fig. 3).

In order to probe the spin-valley physics of the IXs, we measured $\sigma^-$ circularly polarized PL as a function of out-of-plane magnetic field (Figure 3a). In the DC contact region, we measured an exciton g-factor of $7.0 \pm 0.4$ (Figure 3b), consistent with numerous past works on R-type MoSe$_2$-WSe$_2$ heterostructures[1,14,17,20,21]. In the hBN separated region, we measured an exciton g-factor of $-5.4 \pm 1.0$ (Figure 3c), which has not been previously reported. We note that the opposite sign of the g-factor is consistent with our zero field circularly polarized measurements showing that the hBN separated PL has co-circularly polarized PL when pumping at 1.72 eV.

Finally, we compared the temperature dependent behavior of both IX species to investigate the source of localization in the heterostructure. Figures 4a and b shows the temperature dependent PL from both regions of the heterostructure. In Figure 4a, we see the IX emission from the DC region of the heterostructure. With increasing the temperature, the IX emission changes from a series of individual, narrow peaks to become a homogeneous broad peak around 13 K. The behavior of the hBN-separated region shows a disappearance of the narrow IX lines at almost the same temperature (9 K); however, no broad peak persists to higher temperature. In both cases, the narrow lines disappear around 10 K, suggesting that the localization potential that gives rise to the narrow lines is same for both regions and independent of the moiré potential. We note however, that there are differences between the two sample regions. In the DC region, a broad PL peak persists to temperatures above 19 K, whereas in the hBN separated region the PL disappears completely (Figure 4c). See Extended Data Fig. 4 for higher temperatures. The temperature dependent PL of both regions were measured in the same thermal cycle, using the same conditions including exposure time, power and excitation wavelength. The measurement was repeated several times with similar results. We therefore present a simple picture that explains all of the observed behaviors. In the DC region, a weak extrinsic localization potential sits on top of the deep ~50-100 meV moiré potential (Figure 4d). The moiré potential explains the lower IX energy in the DC region, the cross-circularly polarized PL, and the positive g-factor; however, the narrow lines originate from the shallower extrinsic potential which disappears abruptly at ~10 K. However, the broader PL signal persists to higher temperature due to the trapping of IX by the wider, deeper moiré potential. Whereas, on the hBN separated region, the moiré potential is suppressed completely (Figure 4d), explaining the higher IX energy, co-polarized optical selection rule, and negative g-factor. In this region, localization is solely due to the extrinsic potential. As such, the narrow lines are observed below 10 K, but the PL vanishes completely above 19 K since the hBN-separated region does not have the moiré potential to confine the IXs, and the strong dipole-dipole and exchange interaction scatters the IXs outside of the detection area or light cone.

In summary, we have shown that the spectrally narrow IX lines that were previously attributed to intrinsic trapping via the moiré potential are extrinsic in nature and more likely originate from nanoscale defects or nanobubbles formed during the 2D heterostructure fabrication process. In DC heterostructures this extrinsic potential sits on top of the moiré potential giving rise to narrow IX lines that maintain the characteristics of intrinsic moiré IXs. Our result provides crucial insights into future quantum device applications of moiré excitons and motivates the development of improved 2D material fabrication techniques to realize the goal of 2D quantum emitter arrays with homogeneous energies. Our result also demonstrates the requirement of spatially trapping IXs to realize optically detectable PL as evidenced by the disappearance of PL in the hBN separated region as the IXs become delocalized.

## Methods:

### Sample Fabrication
The layers of WSe$_2$, MoSe$_2$, hBN and graphene were exfoliated from bulk using the scotch tape method. The layer thickness was measured by atomic force microscopy and optical contrast. For aligning the TMD monolayers near 0° degree precisely, we used polarization resolved second harmonic generation spectroscopy[27,28]. The device was fabricated using the dry transfer technique[25]. The bottom and top hBN thicknesses were 22 nm and 8 nm respectively. The top gate was bilayer graphene, and the bottom gate was 2 nm thick graphite. 7 nm/ 40 nm chrome gold contacts to the device were patterned using electronic beam lithography and thermal evaporation.

### Optical measurements
For the confocal and polarized PL measurements, we used a 1.72 eV photon energy continuous wave laser (M Squared SOLSTIS) on resonance with WSe$_2$. Unless otherwise noted the excitation power was 20 µW and the sample temperature was 1.6 K. The experiments were performed in the reflection geometry focusing the laser and collecting with a 0.81 NA attocube objective. In the confocal measurements, a detection area of 1 µm was achieved by spatially filtering the PL and using a 50 µm pinhole and a 50× magnification confocal setup. We used appropriate combinations of polarizers and achromatic waveplates to control the excitation and detection polarizations.

### Electronic measurements
The electric field reported in Figure 2a is calculated by $E_{hs} = \left(\frac{V_t - V_b}{t_t + t_b}\right) * \frac{\epsilon_{hBN}}{\epsilon_{hs}}$ where $V_t$ ($V_b$) is the voltage applied to the top (bottom) gate, $t_{t(b)}$ is the thickness of the top (bottom) hBN, $\varepsilon_{hBN} = 3.7$ ($\varepsilon_{hs}$) is the relative dielectric constant of the hBN (heterostructure). The dielectric constant of the heterostructure was determined by taking the weighted average (weighted by layer thickness) of the dielectric constant of the TMD and hBN layers. This is calculated to be $\varepsilon_{hs} = \{(t_{WSe_2} * \varepsilon_{WSe_2}) + (t_{hBN} * \varepsilon_{hBN}) + (t_{MoSe_2} * \varepsilon_{MoSe_2})\}/(t_{WSe_2} + t_{hBN} + t_{MoSe_2}) = 6.26$ where $t_{WSe_2}$, $t_{MoSe_2}$ and $t_{hBN}$ are the thickness of the WSe$_2$, MoSe$_2$ and hBN layers respectively. The doping density is calculated using the parallel plate capacitor model where $\sigma = \varepsilon_{hBN}(V_b + V_t)/t_{total}$.

### DFT calculation
We create a 1 × 1 MoSe$_2$/WSe$_2$ unit cell with a lattice constant of 3.317 Å, which is the average of the MoSe$_2$ and WSe$_2$ lattice constants. All calculations of the **K** point bandgap ($E_g$) of MoSe$_2$/WSe$_2$ hetero-bilayers with different translation **r**$_0$ between layers are performed in the framework of density functional theory by using a plane-wave basis set as implemented in the Vienna ab initio simulation package. For each given **r**$_0$, we fix the interlayer distance $d = 6.447$ Å, which is the minimum value among all given **r**$_0$ in our calculations. We calculate $E_g$ for this interlayer distance along with a series of increasing interlayer distances of 6.547 Å, 6.647 Å, 6.747 Å, 6.847 Å, 6.947 Å, 7.447 Å, and 9.447 Å. These increasing distances simulate the effect of separating the layers with the addition of hBN.

The electron-ion interactions are modeled using the projector augmented wave (PAW) potentials. The generalized gradient approximation (GGA) with the Perdew-Burke-Ernzerhof (PBE) functional with van der Waals corrections (vdW) for the exchange-correlation interactions is used. In all of our calculations the spin-orbit coupling (SOC) is fully taken into account. A vacuum of 15 Å is used in all the calculations to avoid interaction between the neighboring slabs. The plane-wave cutoff energy is set to 500 eV and the first Brillouin zone of the unit cell of $MoSe_2/WSe_2$ hetero-bilayers is sampled by using the Monkhorst-Pack scheme of $k$-points with the $12 \times 12 \times 1$ mesh for the structural optimization and the $24 \times 24 \times 1$ mesh for the band structure. The residual forces have converged to less than 0.01 eV/Å and the total energy difference to less than $10^{-5}$ eV.

**Data availability:**

The data that support the findings of this study are available from the corresponding author upon reasonable request.

**Code availability:**

Upon request, authors will make available any previously unreported computer code or algorithm used to generate results that are reported in the paper and central to its main claims.

**Acknowledgments:**

**General:**

We thank Hongyi Yu who performed the DFT calculations.

**Funding:**

JRS and BJL acknowledge support from the National Science Foundation Grant. Nos. DMR-2003583 and ECCS-2054572. JRS acknowledges support from Air Force Office of Scientific Research Grant Nos. FA9550-20-1-0217 and FA9550-21-1-0219. BJL acknowledges support from the Army Research Office under Grant nos. W911NF-18-1-0420 and W911-NF-20-1-0215.

DGM acknowledges support from the Gordon and Betty Moore Foundation's EPiQS Initiative, Grant GBMF9069. KW and TT acknowledge support from JSPS KAKENHI (Grant Numbers 19H05790, 20H00354 and 21H05233).

**Author Contributions**:

JRS and BJL conceived and supervised the project. DNS fabricated the structures and FM performed the experiments, assisted by DNS and MK. FM analyzed the data with input from DNS, JRS and BJL. MRK and DGM provided and characterized the bulk $MoSe_2$ and $WSe_2$ crystals. TT and KW provided the bulk hBN crystals. FM, DNS, JRS, and BJL wrote the paper with input from all the authors. All authors discussed the results.

**Competing Interests:**

The authors declare no competing interests.

# Figures:

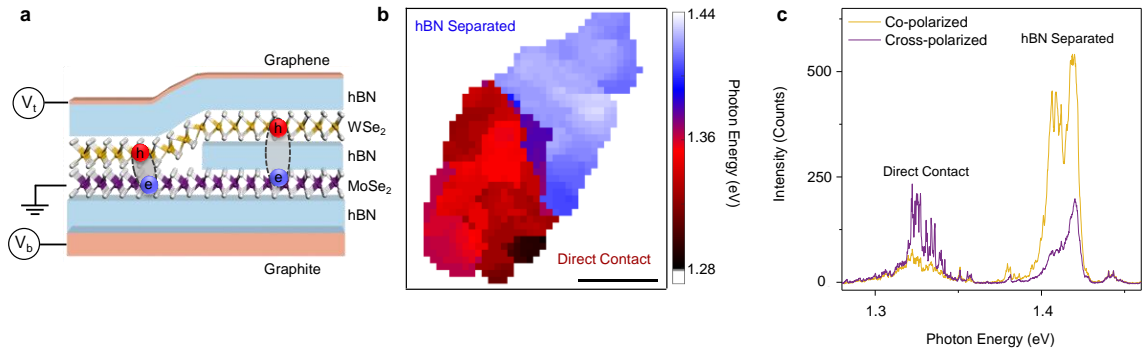

**Fig. 1: Schematic of the sample and spatially resolved IX photoluminescence.**
**a**, Cartoon depiction of the device shows the WSe$_2$-MoSe$_2$ heterostructure encapsulated with hBN. Bilayer graphene is used for the top gate and graphite is used for the bottom gate. Approximately half of the TMD heterostructure is separated by bilayer hBN to suppress the moiré potential. **b**, Confocal PL spatial map of this device, plotting the average center IX photon energy. The hBN separated region is shown in blue with 1.37 to 1.44 eV IX emission energy and the direct contact area is shown in red with the emission energy between 1.28 to 1.36 eV. **c**, Co- and cross-circularly polarized PL spectra measured by exciting the hBN separated region with a polarized 1.72 eV laser and collecting from both regions at the same time. The signal from the hBN separated region ~1.42 eV is mostly co-polarized, whereas the signal from the DC region ~1.33 eV is mostly cross-polarized.

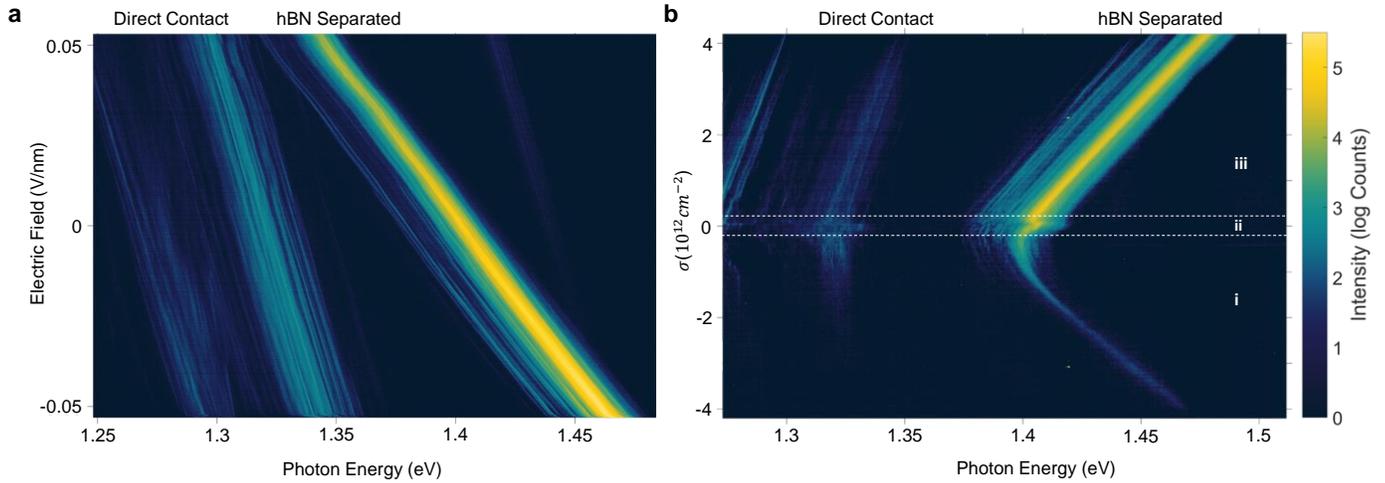

**Fig. 2: IX photoluminescence as a function of electric field and doping level.**
**a,** PL emission as a function of electric field measured by exciting the hBN separated region while collecting from both regions. The hBN separated IX lines show a larger dipole moment due to the increased electron-hole separation. **b**, PL emission as a function of doping level. The regions i, ii, and iii correspond to electron doping, neutral, and hole doping respectively.

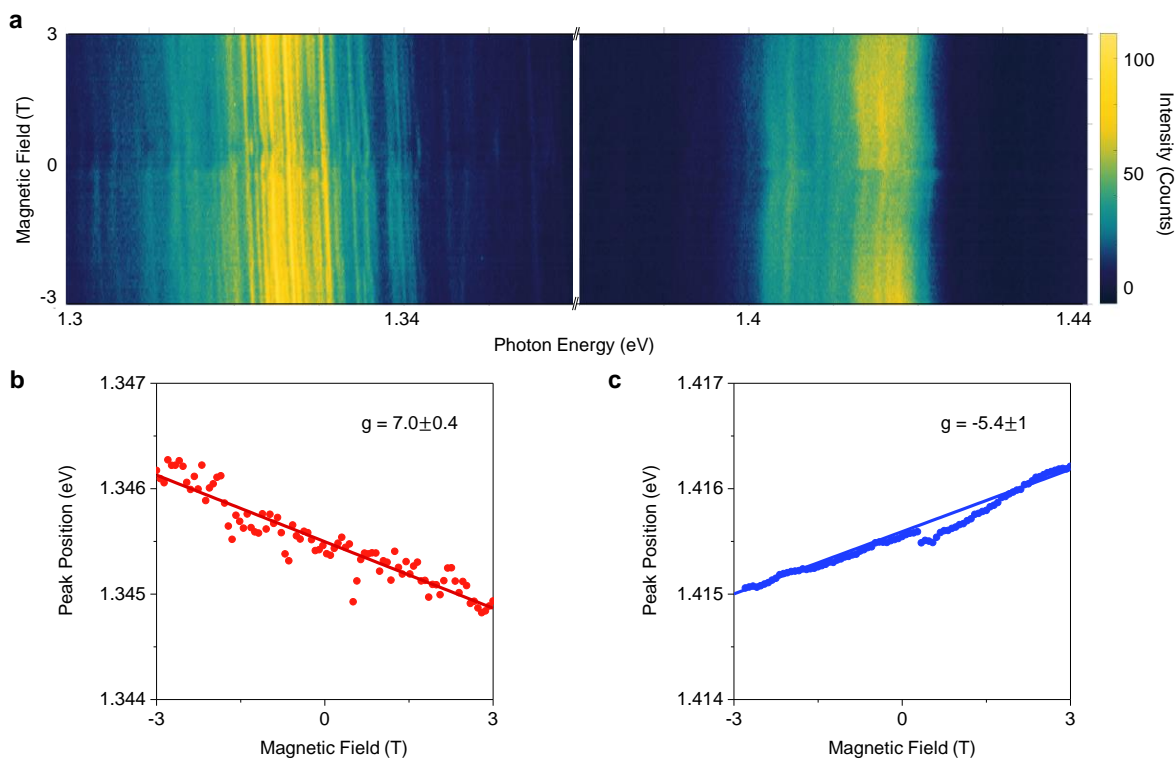

**Fig. 3: Photoluminescence as a function of magnetic field.**
**a,** Magnetic field dependent PL (detecting $\sigma^-$) shows opposite sign of g-factors for the DC and hBN separated regions. **b-c,** Example magnetic field dependence of a single IX line for both DC (**b**) and hBN separated regions (**c**). The excitonic g-factors reported are average values of the fitting to six single IX lines. The error bar shows one standard deviation.

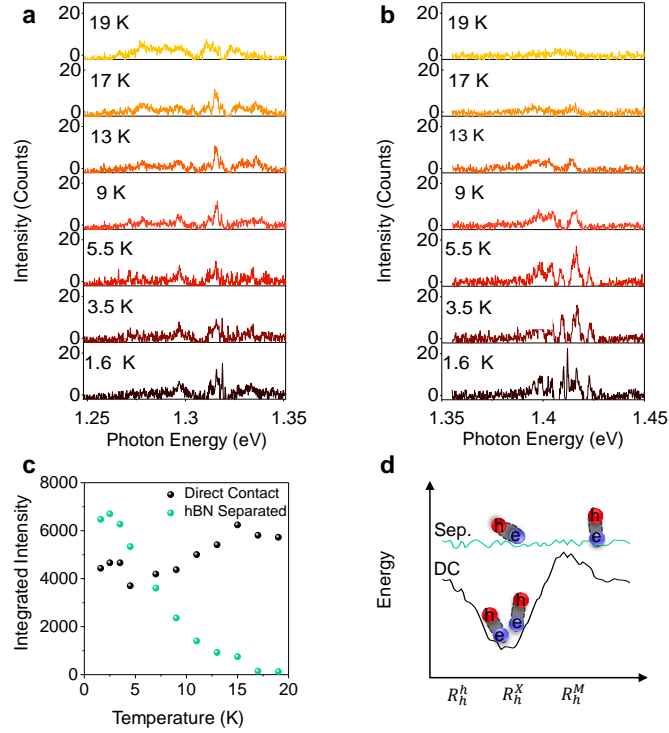

**Fig. 4: Temperature dependent PL and origin of narrow IX lines.**
PL of the narrow IX lines in the DC region as a function of temperature shows the width of individual lines are increasing by 9 K and disappear completely by 17 K. The signal in the DC region includes narrow emission on top of a wider PL plateau. The wider emission persists to higher temperature (see Extended Data Fig.4) that is consistent with previous temperature dependent measurements on R-type heterostructures. **b,** Temperature dependent PL from the hBN separated region shows the narrow lines are widening by 9 K which is in good agreement with the DC region's temperature dependent PL. The hBN separated IX signal disappears fully by 13 K and does not have a wider PL plateau. **c,** Spectrally integrated PL shows IX emission in the hBN separated region disappears completely by 19 K whereas the signal from the DC region is approximately constant. **d,** Depiction of the IX energy as a function of position in the moiré, for DC region (black) and hBN separated (Sep.) region (green). Both regions exhibit a weak extrinsic trapping potential denoted by the fluctuations. The DC region has both moiré trapping and extrinsic fluctuations.

**Extended Data:**

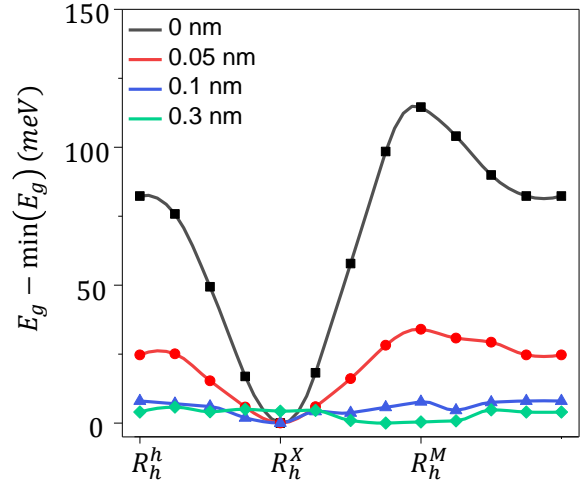

**Extended Data Fig. 1: Interlayer band gap for MoSe$_2$-WSe$_2$ heterostructure.**
Density functional theory simulation of the R-type stacking MoSe$_2$-WSe$_2$ heterostructure shows the moiré potential's change with increasing the vertical distance between layers. $E_g$ is the interlayer band gap, the x-axis labels the high symmetry points of the moiré. The moiré potential disappears when the interlayer distance is increased more than 0.3 nm from the equilibrium separation.

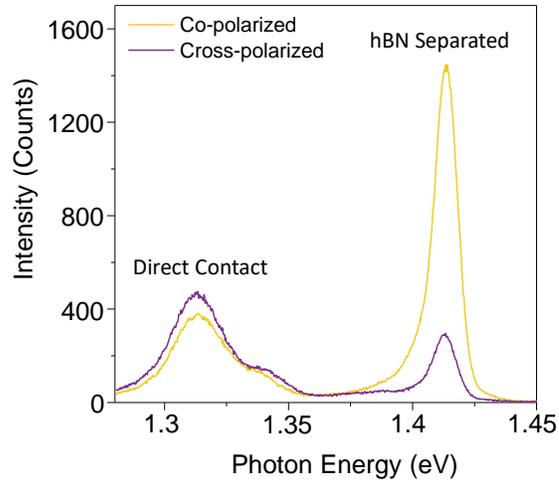

**Extended Data Fig. 2: Direct contact and hBN separated polarization on another sample.**
Circularly polarized PL on another R-type sample. A 1.72 eV laser excited the hBN separated area while PL was collected from both regions. The PL from the hBN separated region centered at 1.41 eV shows co-circularly polarized dominant emission while the DC region's PL at 1.31 eV is cross-circularly polarized. The higher excitation power (20 µW) for this measurement results in the disappearance of the narrow lines and emergence of the wide PL spectrum.

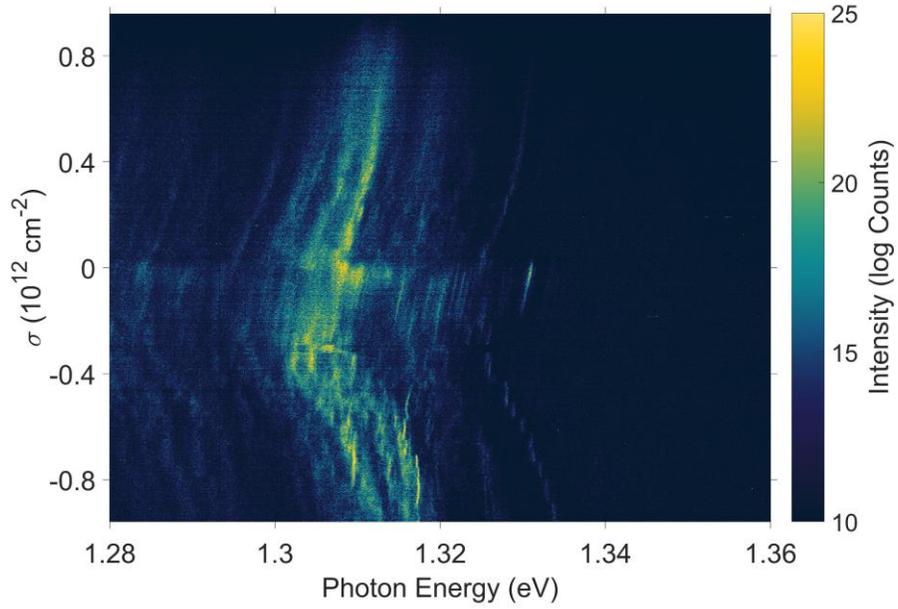

**Extended Data Fig. 3: Fine structure in direct contact region.**
High resolution charge density dependent PL map of the DC region shows fine structure that is attributed to the Coulomb staircase[1]. The high energy line at 1.335 eV is shows ~10 energy steps by increasing the doping of the system from -0.2×10$^{12}$ to -0.9×10$^{12}$ cm$^{-2}$.

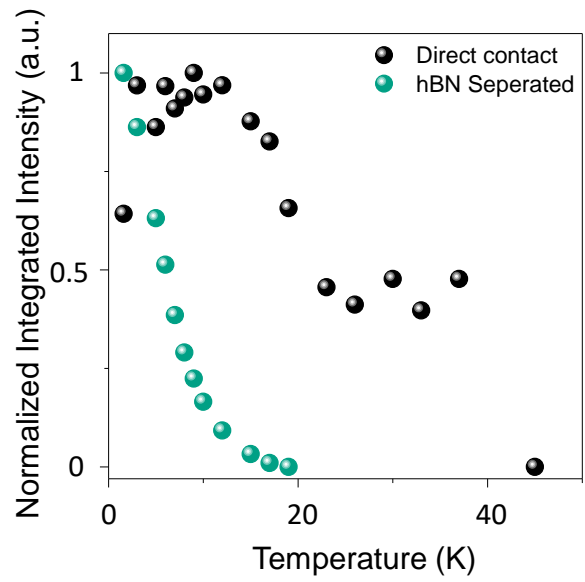

**Extended Data Fig. 4: Temperature dependent PL.**
Higher temperature PL measurement shows that PL in DC region persists to ~40 K.